\documentclass[conference]{IEEEtran}
\IEEEoverridecommandlockouts
\usepackage{cite}
\usepackage{amsmath,amssymb,amsfonts}
\usepackage{algorithmicx}
\usepackage{algpseudocode}
\usepackage[ruled,linesnumbered]{algorithm2e}
\usepackage{hyperref}
\usepackage{graphicx}
\usepackage{textcomp}
\usepackage{booktabs}
\usepackage{subcaption}
\usepackage{xcolor}
\def\BibTeX{{\rm B\kern-.05em{\sc i\kern-.025em b}\kern-.08em
    T\kern-.1667em\lower.7ex\hbox{E}\kern-.125emX}}
\begin{document}
\makeatletter
\newcommand{\rmnum}[1]{\romannumeral #1}
\newcommand{\Rmnum}[1]{\expandafter\@slowromancap\romannumeral #1@}
\makeatother

\title{Enhancing Information Freshness: An AoI Optimized Markov Decision Process}

\author{
\IEEEauthorblockN{
Jingzehua Xu\IEEEauthorrefmark{1},
Yimian Ding\IEEEauthorrefmark{1},
Yiyuan Yang\IEEEauthorrefmark{2},
Guanwen Xie\IEEEauthorrefmark{1}$^{,+}$,
Shuai Zhang\IEEEauthorrefmark{3}$^{,+}$
}
\IEEEauthorblockA{\IEEEauthorrefmark{1}Tsinghua Shenzhen International Graduate School, Tsinghua University, China}
\IEEEauthorblockA{\IEEEauthorrefmark{2}Department of Computer Science, University of Oxford, United Kingdom}
\IEEEauthorblockA{\IEEEauthorrefmark{3}Department of Data Science,
New Jersey Institute of Technology, USA}
Email: sz457@njit.edu

\thanks{$^{+}$ These authors contributed equally to this work.

$^{1}$$\,$ The source code associated with this article is available at the following GitHub repository: https://github.com/Xiboxtg/AoI-MDP}
}

\maketitle

\begin{abstract}
Ocean exploration utilizing autonomous underwater vehicles (AUVs) via reinforcement learning (RL) has emerged as a significant research focus. However, underwater tasks have mostly failed due to the observation delay caused by information limitation in the information updating networks. In this study, we present an AoI optimized Markov decision process (AoI-MDP) to improve the performance of underwater tasks. Specifically, AoI-MDP models observation delay as timing delay through statistical delay formulation, and includes this delay as a new component in the state space. Additionally, we introduce wait time in the action space, and integrate AoI with reward functions to achieve joint optimization of information freshness and decision-making for AUVs leveraging RL for training. Finally, we apply this approach to the multi-AUV data collection task scenario as an example. Simulation results highlight the feasibility of AoI-MDP, which effectively minimizes AoI while showcasing superior performance in the task. To accelerate relevant research in this field, we have made the simulation codes available as open-source$^{1}$.
\end{abstract}


\section{Introduction}
Harsh ocean environment put forward higher difficulty on ocean exploration \cite{1}. As a novel approach, utilizing autonomous underwater vehicles (AUVs) via reinforcement learning (RL) has merged as a significant research focus \cite{2}. Relying on information updating networks (IUNs) \cite{8}, AUVs can communicate with each other and work in collaboration to accomplish human-insurmountable tasks \cite{3}. However, underwater tasks have mostly failed due to the observation delay caused by information limitation, leading to the non-causality of control policies \cite{4}. Although this issue can be alleviated by introducing states that incorporate past information and account for the future effects of control laws \cite{4}, it becomes increasingly challenging as the number of AUVs grows, leading to more complexity in both communication and decision-making processes \cite{5}.

As a significant indicator evaluation the freshness of information, age of information (AoI) is proposed to measure the time elapsed at the receiver since the last information was generated until the most recent information is received \cite{6}. And it has been verified to solve the severe delay caused by constantly sampling and transmitting observation information \cite{7}. Central to this consensus is that minimizing AoI can enhance the freshness of information, thereby facilitating efficiency of subsequent decision-making process in the presence of observation delay \cite{7}. Currently, numerous studies have focused on optimizing AoI to aid decision-making in the context of land-based or underwater tasks. For example, Messaoudi \textit{et al.} optimized vehicle trajectories relying on minimizing average AoI while reducing energy consumption \cite{10279435}. Similarly, Lyu \textit{et al.} leveraged AoI to assess transmission delay impacts on state estimation, improving performance under energy constraints \cite{9442098}. These studies primarily aim to reduce AoI by improving motion strategies of agents, without considering the impact of information update strategies on AoI. They assume that agents instantaneously perform the current action upon receiving previous information. However, this zero-wait strategy has been shown to be suboptimal in scenarios with high variability in delay times \cite{8000687}. Conversely, it has been demonstrated that introducing waiting time before updating can achieve lower average AoI. This highlights the necessity of integrating optimized information update strategies into underwater tasks. 

Furthermore, most studies currently leverage the standard Markov decision process (MDP) without observation delay to model the underwater tasks, which assumes the AUV can instantaneously receive current state information without delay, so that it can make corresponding actions \cite{Howard1960DynamicPA}. This idealization, however, may not hold in many practical scenarios, since information delay effects and high update frequencies causing channel limitation reduce the freshness of received information, hindering the AUV’s decision-making efficiency. Therefore, extending the standard MDP framework to incorporate observation delays and AoI is necessary \cite{10.1145/149439.133106}.

Based on the above analysis, we attempted to propose an AoI optimized MDP (AoI-MDP) dedicated in the underwater task to improve the performance of the tasks with observation delay. The contributions of this paper include the following:
\begin{itemize}
\item To the best of our knowledge, we are the first to formulate the underwater task as an MDP that incorporates observation delay and AoI. Based on AoI-MDP, we utilize RL for AUV training to realize joint optimization of both information updating and decision-making strategies.

\item Instead of simply modeling observation delay as a random distribution or stationary stochastic variable, we utilize statistical delay formulation to realize the delay-oriented modeling via sensor-based model, which potentially yielding more realistic results.

\item Through comprehensive evaluations and ablation experiments in the underwater data collection task, our AoI-MDP showcases superior feasibility and excellent performance in balancing multi-objective optimization. And to accelerate relevant research in this field, the code for simulation will be released as open-source in the future.

\end{itemize}

\section{Methodology}
In this section, we present the proposed AoI-MDP, which consists of three main components: an observation delay-aware state space, an action space that incorporates wait time, and reward functions based on AoI. To achieve delay-oriented modeling, AoI-MDP utilizes statistical delay formulation (SDF) to represent observation delay as delayed observation timing delay, thereby aiming to minimize the gap between simulation and real-world underwater tasks.
\subsection{AoI Optimized Markov Decision Process}
As illustrated in Fig. 1, consider the scenario where the $i$-th delayed observation signal is transmitted from the AUV at time $T_i$, and the corresponding observed information is received at time $D_i$, AoI is defined using a sawtooth piecewise function
\begin{equation}\Delta(t)=t-T_i,D_i\leq t<D_{i+1}, \forall i \in \mathbf{N}.\end{equation}

Hence, we denote the MDP that integrates observation delay and characterizes the freshness of information through the AoI as the AoI-MDP, which can be defined by a quintuple $\Omega$ for further RL training \cite{9} 

\begin{figure}[!t]
    \centering
    \includegraphics[width=1.0\linewidth]{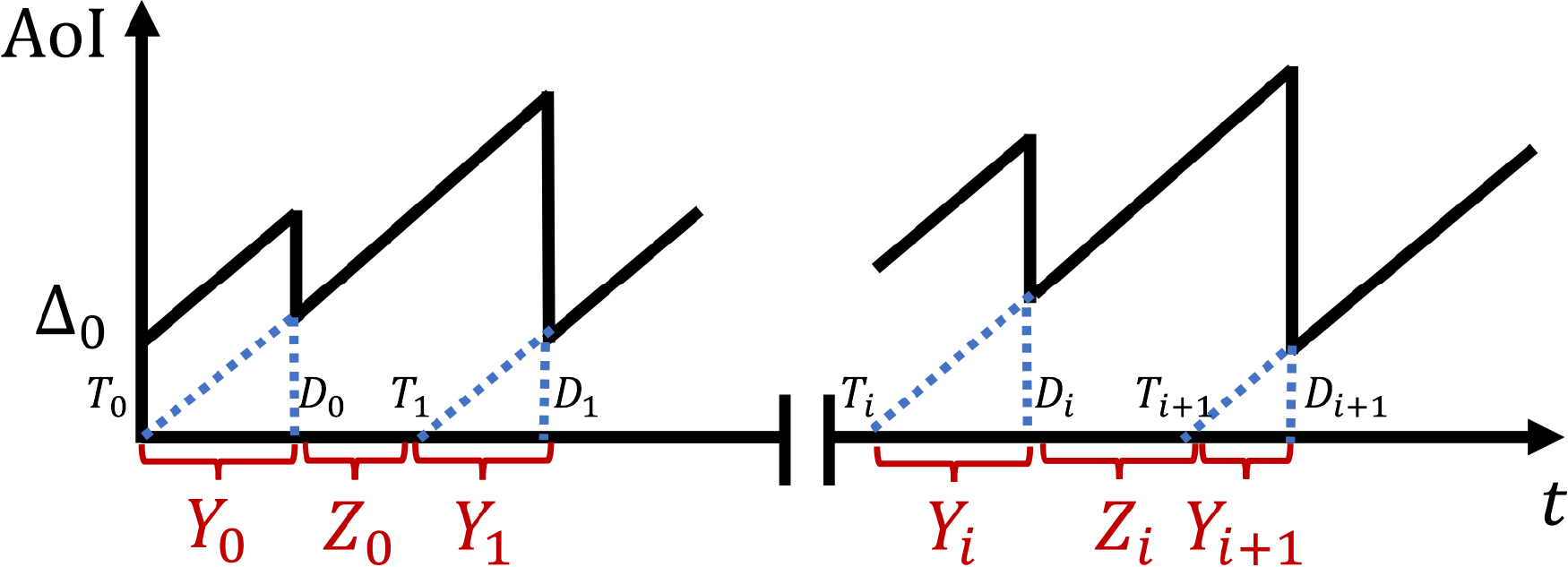}
    \caption{Illustration of the AoI model, which is defined
using a sawtooth piecewise function, where $Y_i$ and $Z_i$ denote the observation delay and wait time at time $i$, respectively.}
    \label{fig_1}
    \end{figure}

\begin{equation}
\Omega=\left \{ \boldsymbol{\mathcal{S}}, \boldsymbol{\mathcal{A}}, \boldsymbol{\mathcal{R}}, {\rm Pr}(s_{i+1}|s_{i}, a_{i}), \gamma\right \} ,
\end{equation}
where $\boldsymbol{\mathcal{S}}, \boldsymbol{\mathcal{A}}, \boldsymbol{\mathcal{R}}$ represent state space, action space and reward functions, respectively. The term ${\rm Pr}(s_{i+1}|s_{i}, a_{i})$ $\in$ [0, 1] indicates state transition probability distribution, while $\gamma$ $\in$ [0, 1] represents a discount factor.

In AoI-MDP, instead of simply incorporating AoI as a component of the reward functions to guide objective optimization through RL training, we also leverage AoI as crucial side information to facilitate decision making. Specifically, we reformulate the standard MDP's state space, action space, and reward functions. The detailed designs for each of these elements are as follows:

\textbf{State Space $\boldsymbol{\mathcal{S}}$:} the state space of the AoI-MDP consists of two parts:  AUV's observed information $s^{'}_i$, and observation delay $Y_i$ at time $i$, represented by $s_i=(s^{'}_i,Y_i)\in \boldsymbol{\mathcal{S^{'}}}\times\boldsymbol{Y}$.We introduce the observation delay $Y_i$ as a new element so that the AUV can be aware of the delayed observation timing delay when the model emits an delayed observation signal to detect the surrounding environment. Additionally, we achieve delay-oriented modeling of both $s^{'}_i$ and $Y_i$ through SDF, whose details are presented in Section \Rmnum{2}-B. 

\textbf{Action Space $\boldsymbol{\mathcal{A}}$:} the action space of the AoI-MDP consists of the tuple $a_i=(a_i^{'}, Z_i)\in \boldsymbol{\mathcal{{A}^{'}}}\times\boldsymbol{Z}$, where $a_i^{'}$ denotes the actions taken by the AUV, while $Z_i$ indicates the wait time between observing the environmental information and decision-making at time $i$. Through jointly optimizing the wait time $Z_i$ and action $a_i$, we aim to minimize the AoI, enabling the AUV's decision-making policy to converge to an optimal level.

\textbf{Reward Function $\boldsymbol{\mathcal{R}}$:} the reward function $r_i^{'}$ in standard MDP comprises elements with different roles, such as penalizing failures, promoting efficiency, and encouraging cooperation, etc. Here, we introduce the time-averaged AoI as a new component of the reward function. Thus, the updated reward function can be represented by the tuple $r_i=(r^{'}_i, -\Bar{\Delta})$. And the time-averaged AoI can be computed as follows:
\begin{equation}
    \Bar{\Delta}=\frac{\sum_{i=1}^{\mathcal{N}}((2Y_{i-1}+Y_i+Z_{i-1})\times(Y_i+Z_{i-1}))+S_0}{2\times(\sum_{i=1}^{\mathcal{N}}Z_{i-1}+\sum_{i=1}^{\mathcal{N}}Y_{i}+Y_0)},
\end{equation}
where $\mathcal{N}$ is the length of information signal, $S_0=0.5 \times (2\Delta_0+Y_0) \times Y_0$. Therefore, the time averaged AoI can be minimized through RL training.

According to the above analysis, the total reward function is set below:
\begin{equation}R_i=\sum_{k=1}^\infty\lambda^{(k)}r^{(k)}_i,\end{equation}
where $\lambda^{(k)}$ represents the weighting coefficient of the $k$-th reward function.

Based on proposed AoI-MDP, we further integrate it with RL training for the joint optimization of
information freshness and decision-making for AUVs. The pseudocode for the AoI-MDP based RL training is showcased in Algorithm 1.

\begin{algorithm}[!t]
\label{alg:1}
\caption{AoI-MDP Based RL Training}
Initialize the replay buffer $\mathcal{D}$, critic network, and actor network parameters of each AUV.

\For{each epoch $k$}{
Reset the training environment and parameters.

\For{each time step $i$}{

\For{each AUV $j$}{
Obtain current state $s_i^{'}$ and observation delay $Y_i$.

Sample action $a_i^{'}$ and waiting time $Z_i$ according to the actor network.

Wait $Z_i$ and execute $a_i^{'}$ while receiving reward $R_i$.

\While{In delay period}{
Extract $N$ tuples of data ${(\boldsymbol{s}_n,\!a_n\!,R_n,\boldsymbol{s}_{n+1})}_{n\!=\! 1,\!\cdots\!,N}$ \!from $\mathcal{D}$.

Update the Critic Nework.

Update the Actor Network.

}

Store $(\boldsymbol{s}_i,a_i,R_i,\boldsymbol{s}_{i+1})$ in $\mathcal{D}$.

\While{In waiting period}{Repeat the process in delay period.}
}
}}
\end{algorithm}

\subsection{Observation Delay and Information Modeling}
Different from previous work, our study enhances the state space of AoI-MDP by considering the observed information using estimated information perceived by AUV-equipped sensors. And we consider observation delay as delayed observation timing delay, rather than merely treating it as a random distribution \cite{8000687} or stationary stochastic variable \cite{10,11}. This approach aims to provide delay-oriented modeling to improve the performance in the underwater environment. And the schematic diagram is shown in Fig. 2.

\begin{figure}[!t]
    \centering
    \includegraphics[width=1.0\linewidth]{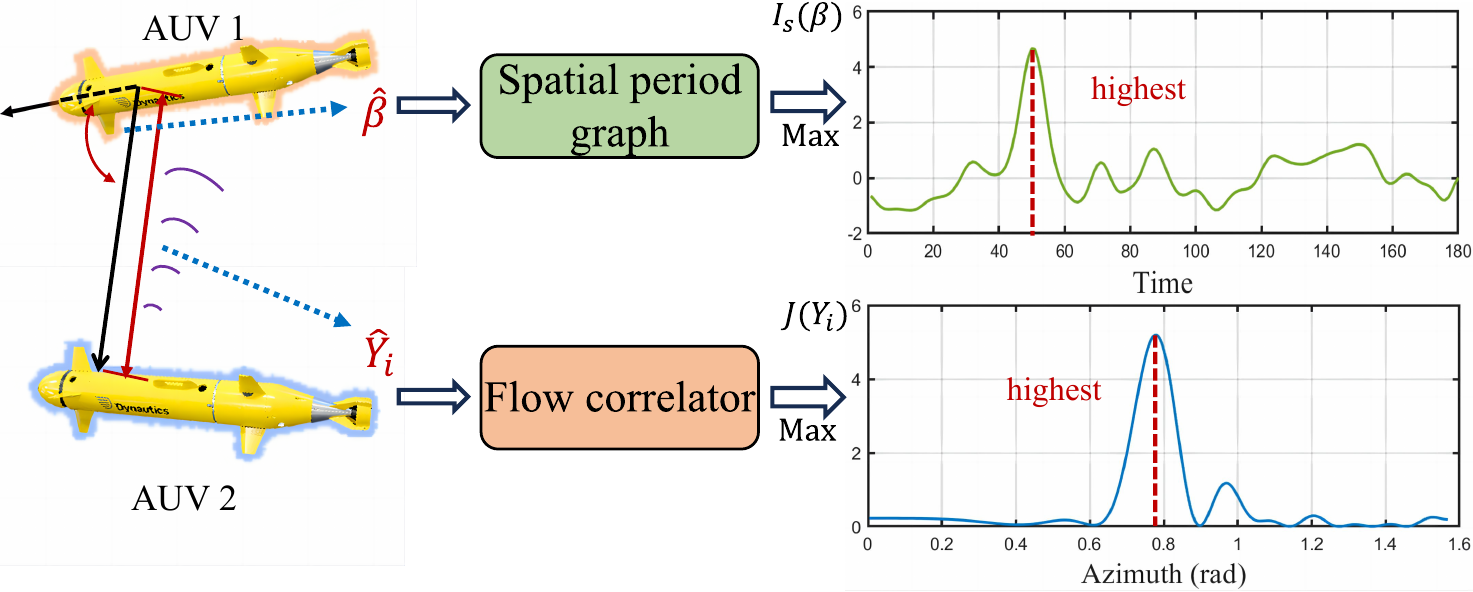}
    \caption{Illustration of the heading and time delay estimation.}
    \label{fig_2}
\end{figure}

To be specific, our study assumes the AUV leverages a sensor model to estimate the distance from itself to environmental objects. This was achieved by transmitting observed signals through sensor, modeling the time delay taken for these signals to propagate to the target, reflect, and return to the sensor hub, thus allowing for distance estimation. The observed signal propagation can be represented as
\begin{equation}\mathcal{X}[n]=\mathcal{S}[n-Y_i]+\mathcal{W}[n], n=0,1,\dots,N-1,\end{equation}
where $\mathcal{S}[n]$ represents the known series, while $Y_i$ denotes the time delay to be estimated, and $\mathcal{W}[n]$ is the Gaussian white noise with variance $\sigma^2$.

We further employ the delay estimator as an estimator to determine the time delay. Specifically, this estimator carries out the following computations on each received signal:
\begin{subequations}
\begin{align}
J[Y_i]=\sum_{n={Y_i}}^{Y_i+M-1}&\mathcal{X}[n]\mathcal{S}[n-Y_i], 0\leq Y_i\leq N-M,\\
&\hat{Y_i} = {\rm argmax} \left[J[Y_i] \right],
\end{align}
\end{subequations}
where $M$ is the sampling length of $\mathcal{S}[n]$. By calculating the value of $\hat{Y_i}$ that maximizes the value of Eq. (6a), the estimated time delay value can be obtained through Eq. (6b).

On the other hand, the AUV in our study utilizes a generic sensing approach to estimate the heading $\beta$ between its orientation and environmental objects. The signal propagation can be expressed as follows:

\begin{equation}
\begin{aligned}
x \left[n\right] = & A c o s \left[2 \pi \left(F_{0} \frac{d}{c} c o s \beta\right) n + \phi\right] + \mathcal{W}[n] , \textrm{ } \\
& n = 0,1 , \cdot \cdot \cdot , M - 1,
\end{aligned}
\end{equation}
where $F_0$ denotes the frequency of observation signal, while $d$ represents the interval between sensors. Besides, $c$ indicates the speed of delayed observation signal propagation, while $A$ and $\phi$ are unknown series amplitude and phase, respectively.

The estimator in SDF is further leveraged to estimate the heading $\beta$. By maximizing the spatial period graph, the estimate of $\beta$  (0\textless$\beta$\textless$\pi$/2) can be calculated

\begin{subequations}
\begin{align}
\!\!\!I_{s} \!\left(\beta\right) \!\!=\!\! \frac{1}{M} &\left(\!\left|\right. \sum_{n = 0}^{M - 1} x \left[n\right] {\rm exp} \left[\right.\! -\! j 2 \pi \!\left(\right.\! F_{0} \frac{d}{c} c o s \beta\! \left.\right)\!n \left]\right. \left|\right.\right)^{2}\!\!,\\
&\quad\quad\quad\quad\hat{\beta} = {\rm argmax} \left[I_{s} \left(\beta\right)\right].
\end{align}
\end{subequations}

By calculating the value of $\beta$ that maximizes the value of Eq. (8a), the estimated time delay value can be obtained through Eq. (8b).

Finally, the AUV can achieve object positioning using the observed $\hat{Y_i}$ and $\hat{\beta}$. These observations are then utilized as data for the observed information in the state space of the AoI-MDP, which potentially
yields more realistic results to improve the underwater performance, while reducing the gap between simulation and reality in the underwater tasks.

\section{Experiments}
In this section, we validate the proposed AoI-MDP through extensive simulation experiments. Further, we present the experimental results with further analysis and discussion.
\subsection{Task Description and Settings}
Since open-source\! underwater\! tasks are scarce, we consider the scenario of a multi-AUV data collection task as a classic example to evaluate the feasibility and effectiveness of the AoI-MDP. This task utilizes RL algorithms to train AUVs to collect data of sensor nodes in the information updating networks, encompassing multiple objectives, such as maximizing sum info rate and collision avoidance, while minimizing energy consumption, etc. For the remaining details and parameters of the task, please refer to the previous work \cite{3}.

\subsection{Experiment Results and Analysis}
We first compared the experimental results of RL training based on AoI-MDP and standard MDP under identical conditions, respectively. Results in Fig. 3 show that AoI-MDP results in lower time-averaged AoI, reduced energy consumption, higher sum info rate, and greater cumulative rewards. This demonstrates that AoI-MDP improves the training effectiveness and performance of the RL algorithm.

% \begin{figure}[!t]

% \begin{minipage}[b]{0.48\linewidth}
%   \centering
%   \centerline{\includegraphics[width=4.0cm]{AoI_wait_or_zero_new.pdf}}
% %  \vspace{2.0cm}
%   \centerline{(a) Average AoI}\medskip
% \end{minipage}
% %
% \begin{minipage}[b]{0.47\linewidth}
%   \centering
%   \centerline{\includegraphics[width=4.0cm]{ec_wait_or_zero_new.pdf}}
% %  \vspace{1.5cm}
%   \centerline{(b) Energy comsumption}\medskip
% \end{minipage}
% \hfill
% \begin{minipage}[b]{0.488\linewidth}
%   \centering
%   \centerline{\includegraphics[width=4.0cm]{smr_wait_or_zero_new.pdf}}
% %  \vspace{1.5cm}
%   \centerline{(c) Sum info rate}\medskip
% \end{minipage}
% %
% \begin{minipage}[b]{0.488\linewidth}
%   \centering
%   \centerline{\includegraphics[width=4.0cm]{reward_wait_or_zero_new.pdf}}
% %  \vspace{1.5cm}
%   \centerline{(d) Cumulative reward}\medskip
% \end{minipage}
% %
% \caption{Comparison of experimental results of algorithms based on AoI-MDP and standard MDP.}
% \label{fig:1}
% %
% \end{figure}

\begin{figure}[!t]
    \centering
    \begin{subfigure}{0.24\textwidth}
        \centering
        \includegraphics[width=\linewidth]{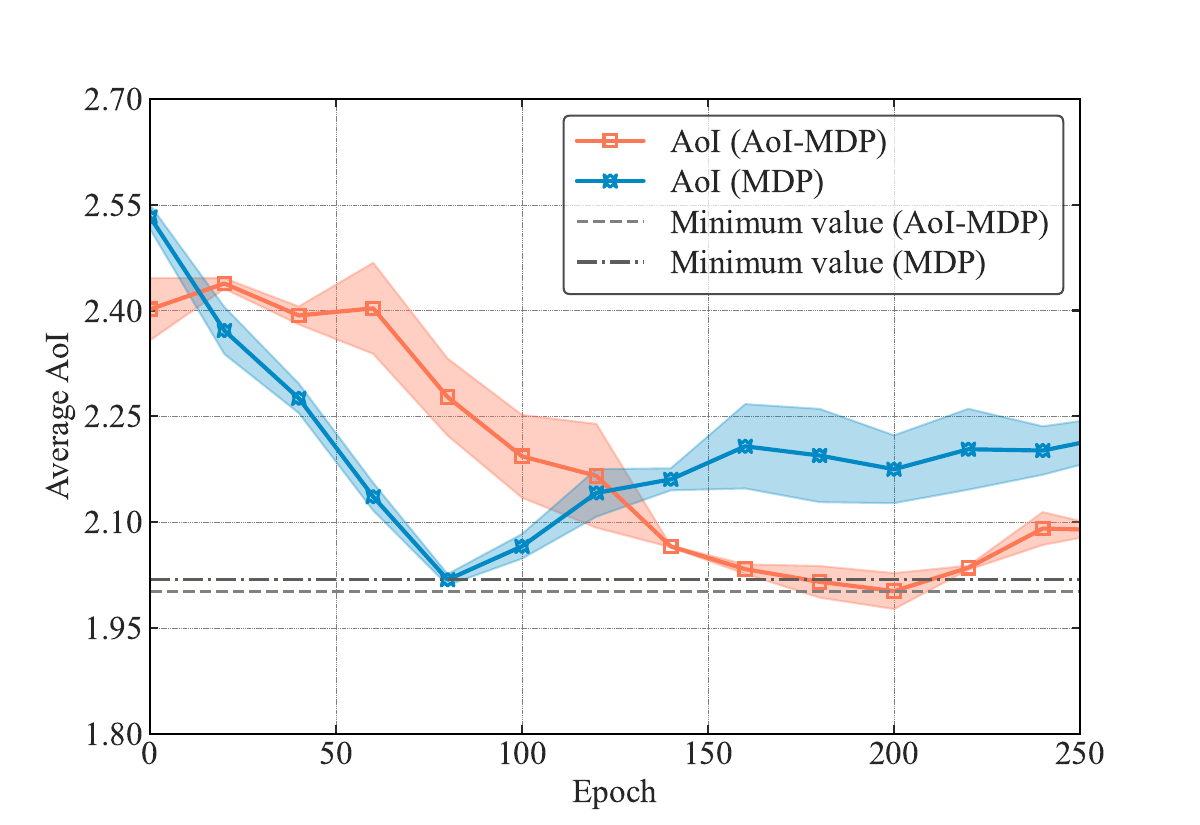}
        \caption{Average AoI}
        \label{fig:subfig1}
    \end{subfigure}
    \hfill
    \begin{subfigure}{0.24\textwidth}
        \centering
        \includegraphics[width=\linewidth]{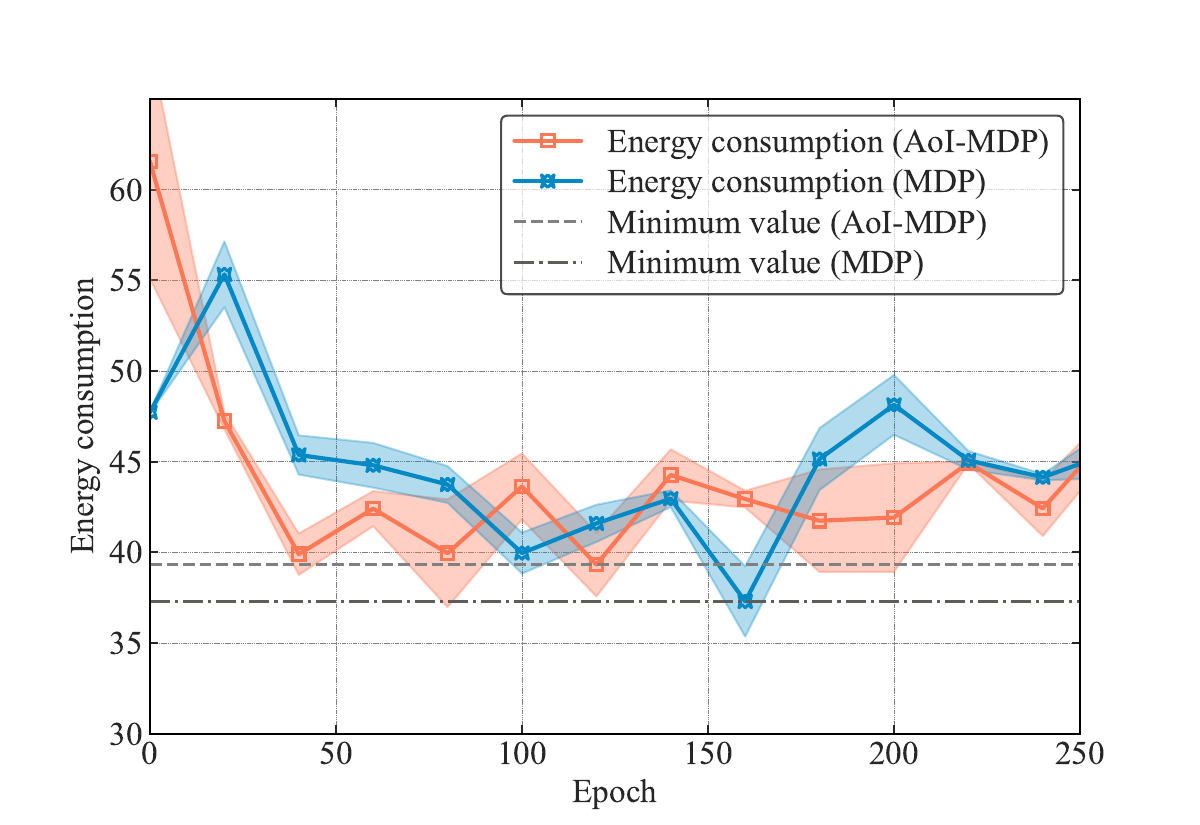}
        \caption{Energy consumption}
        \label{fig:subfig2}
    \end{subfigure}
    
    % \vspace{0.5em}
    
    \begin{subfigure}{0.24\textwidth}
        \centering
        \includegraphics[width=\linewidth]{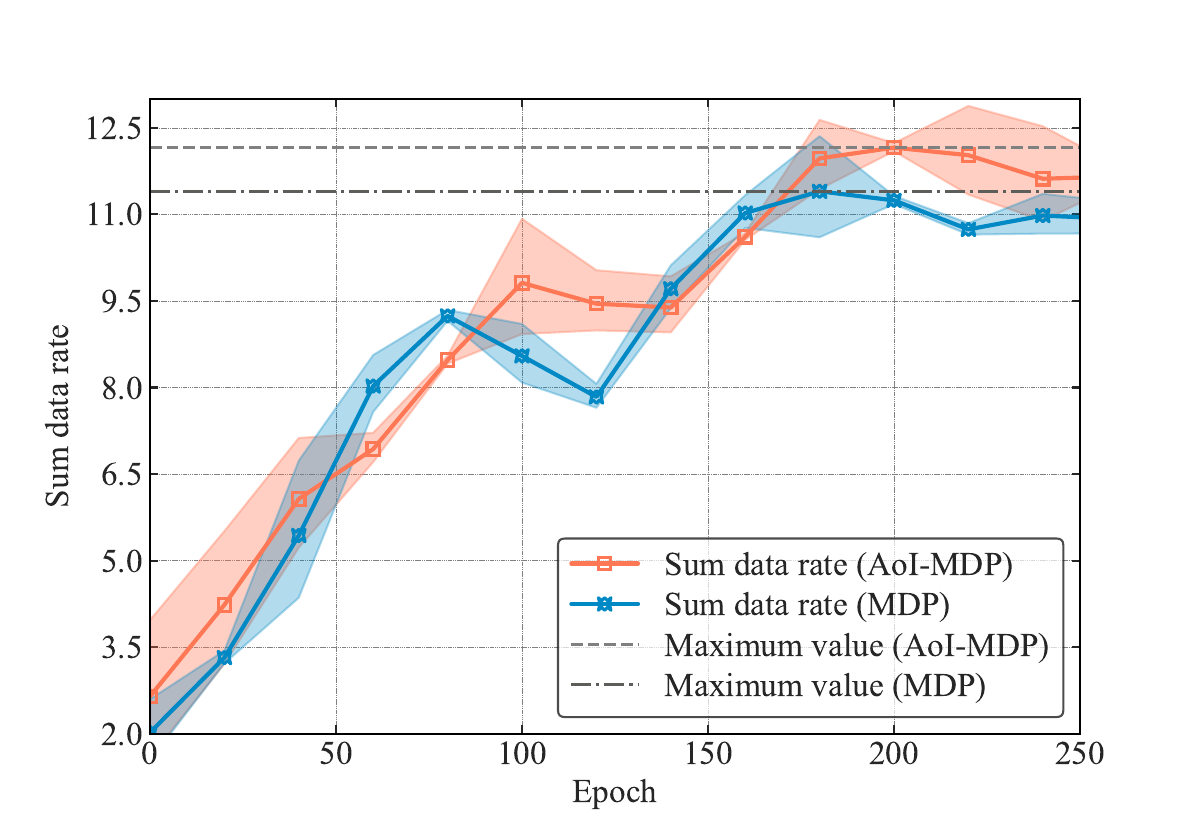}
        \caption{Sum info rate}
        \label{fig:subfig3}
    \end{subfigure}
    \hfill
    \begin{subfigure}{0.24\textwidth}
        \centering
        \includegraphics[width=\linewidth]{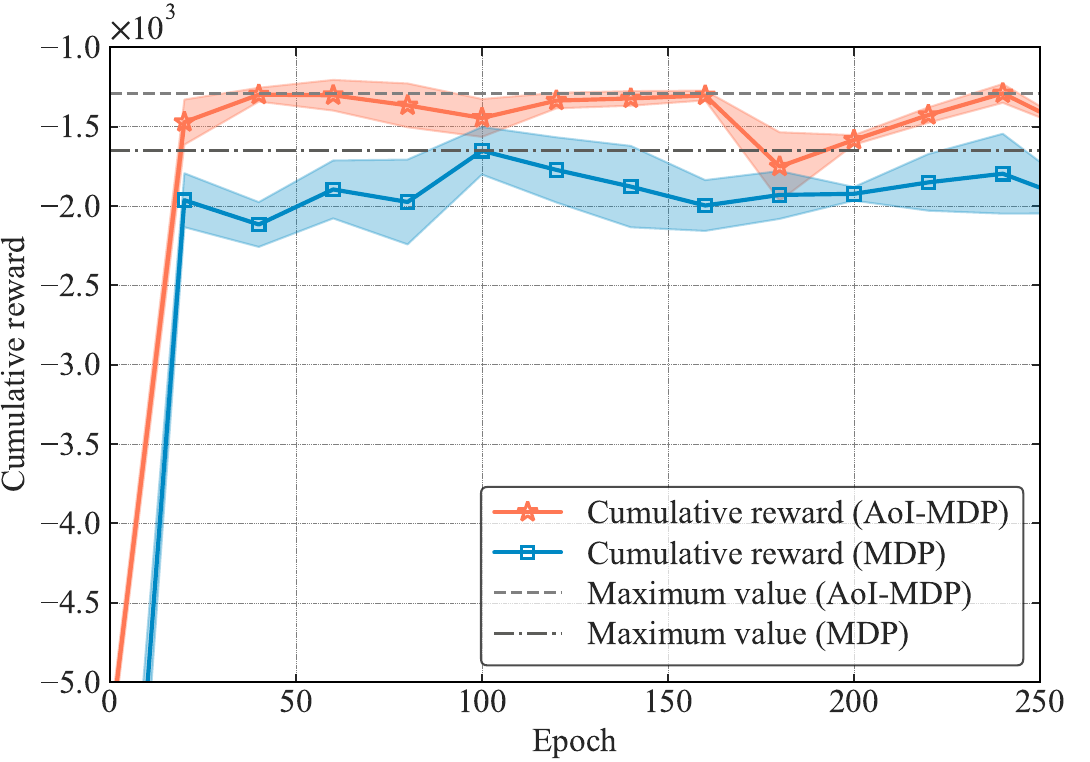}
        \caption{Cumulative reward}
        \label{fig:subfig4}
    \end{subfigure}
    
    \caption{Comparison of experimental results of RL training based on AoI-MDP and standard MDP.}
    \label{fig:overall}
\end{figure}

\begin{table}[!t]
    \centering
    \caption{Comparison of different delay models.}
    \label{table:masking_performance}
    \begin{tabular}{lcccc}
    \toprule
      & AoI & Sum info rate & Energy consumption  \\
    \midrule
    SDF          & 1.97±0.26 & 11.99±0.73 & 33.83±2.59 \\
    Poisson      & 3.42±0.18  & 5.95±2.42  & 34.27±7.98  \\
    Exponential  & 2.67±0.26   & 7.65±1.99  & 43.11±4.16  \\
    Geometric    & 2.38±0.28  & 12.34±0.79 & 58.15±9.49 \\
    \bottomrule
    \end{tabular}
\end{table}

Then we evaluated the generalization performance of AoI-MDP using commonly employed delay models in the communication field, including exponential, poisson and geometric distributions. The experimental results, compared with the SDF model, are shown in Table 1. The AoI-MDP based RL training demonstrates superior performance across various distributions, indicating strong generalization capabilities. Additionally, SDF for time delay modeling achieved near-optimal results in AoI optimization, sum info rate optimization, and energy consumption optimization, underscoring its effectiveness in the underwater data collection task.

We further turned our attention to comparing the generalization of AoI-MDP in various RL algorithms. We conducted experiments utilizing AoI-MDP on soft actor-critic (SAC) and conservative Q-learning (CQL), within the contexts of online and offline RL, respectively. As shown in Fig. 4, both online and offline RL algorithms can successfully adapt to AoI-MDP, while ultimately achieving favorable training outcomes.

\begin{figure}[!htbp]
    \centering
    \begin{subfigure}{0.240\textwidth}
        \centering
        \includegraphics[width=\linewidth]{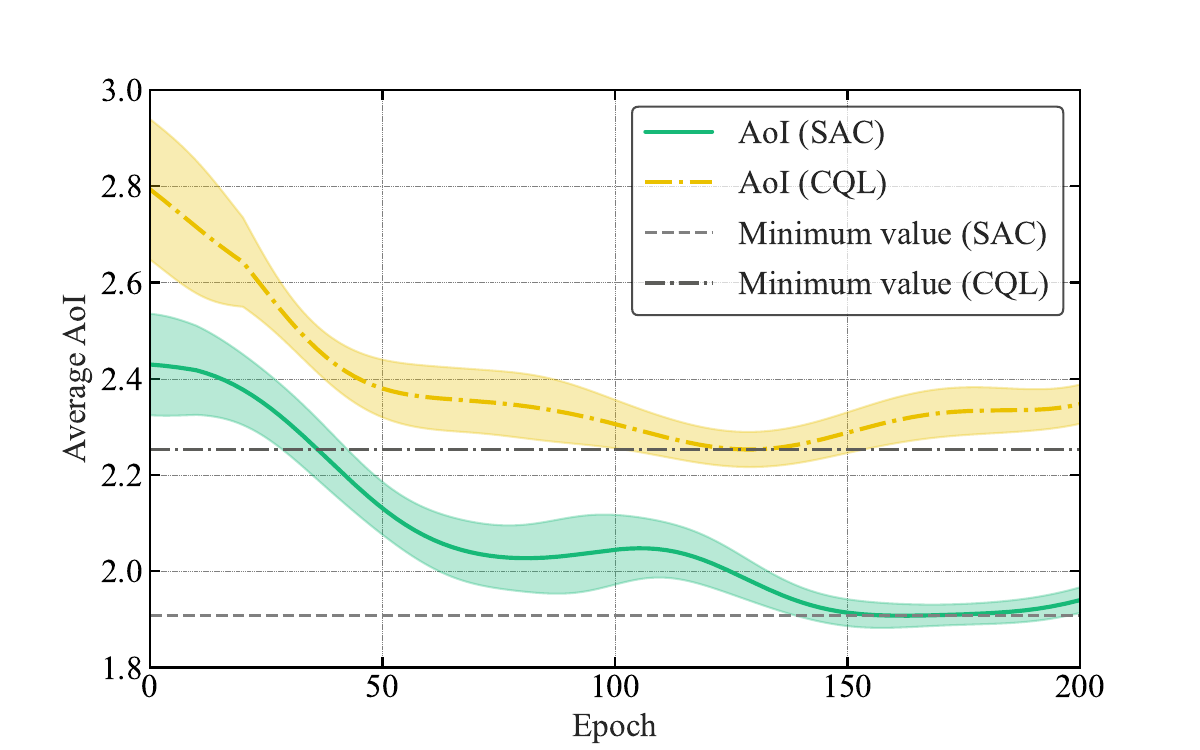}
        \caption{Average AoI}
        \label{fig:subfig1}
    \end{subfigure}
    \hfill
    \begin{subfigure}{0.240\textwidth}
        \centering
        \includegraphics[width=\linewidth]{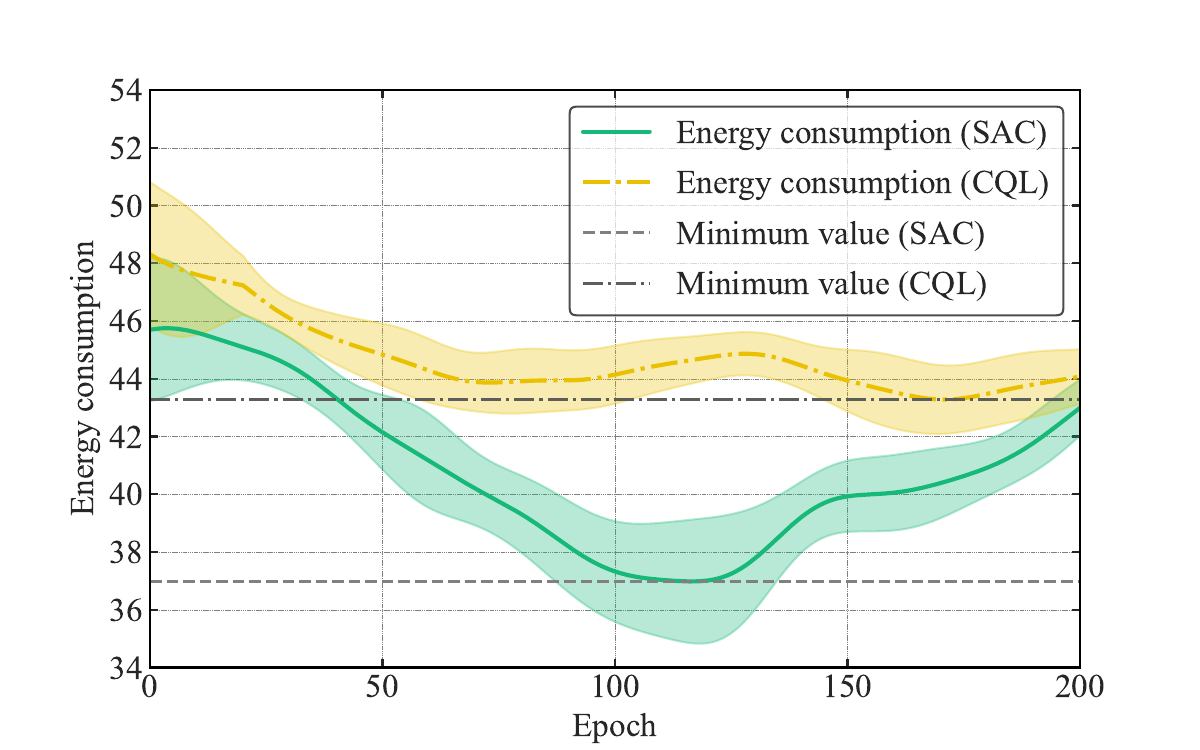}
        \caption{Energy consumption}
        \label{fig:subfig2}
    \end{subfigure}
    
    % \vspace{0.5em}
    
    \begin{subfigure}{0.240\textwidth}
        \centering
        \includegraphics[width=\linewidth]{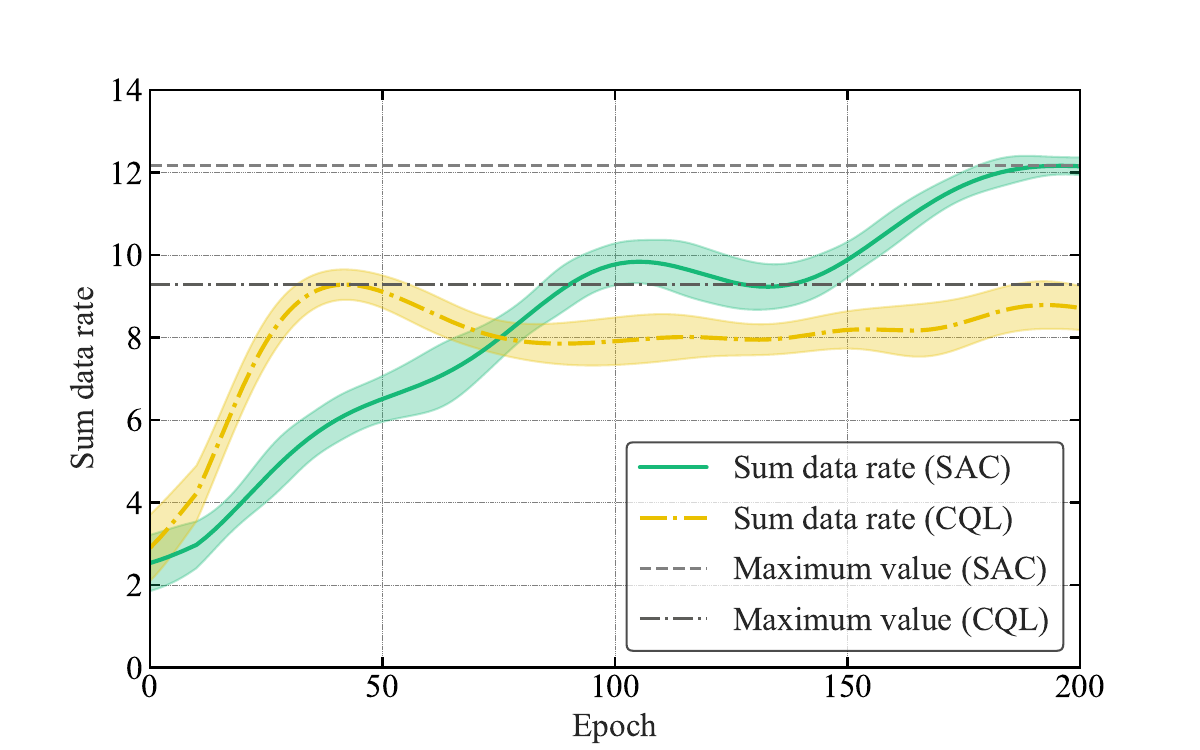}
        \caption{Sum info rate}
        \label{fig:subfig3}
    \end{subfigure}
    \hfill
    \begin{subfigure}{0.240\textwidth}
        \centering
        \includegraphics[width=\linewidth]{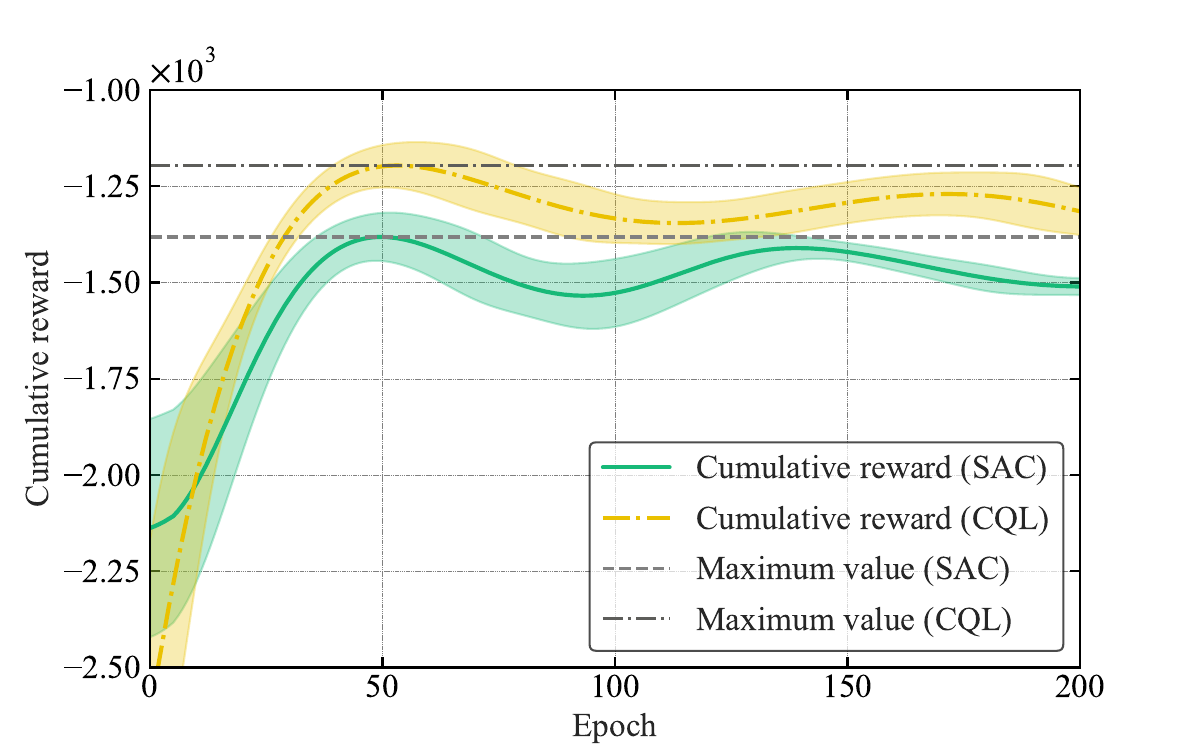}
        \caption{Cumulative reward}
        \label{fig:subfig4}
    \end{subfigure}
    
    \caption{Comparison of experimental results using online and offline RL algorithms based on AoI-MDP.}
    \label{fig:overall}
\end{figure}

\begin{figure}[!t]

\begin{minipage}[b]{0.49\linewidth}
  \centering
  \centerline{\includegraphics[width=4.0cm]{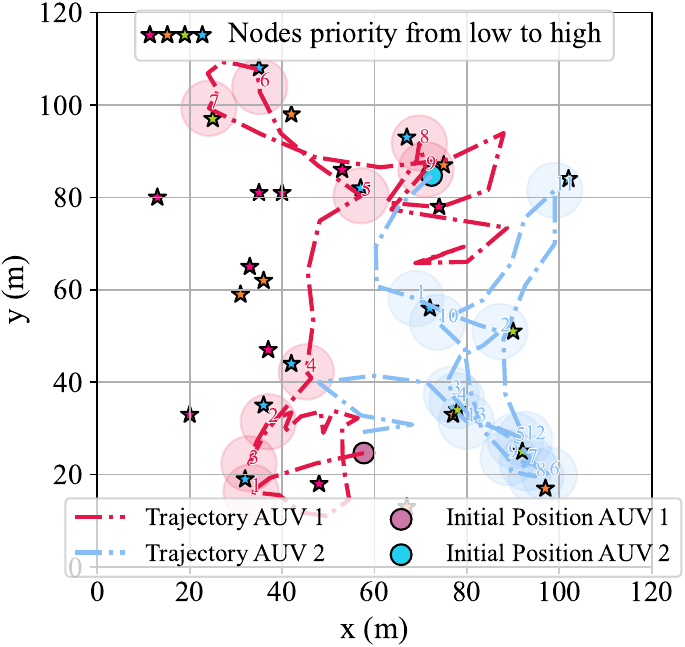}}
%  \vspace{2.0cm}
  \centerline{(a) AUV trajectories (AoI-MDP)}\medskip
\end{minipage}
\begin{minipage}[b]{0.49\linewidth}
  \centering
  \centerline{\includegraphics[width=4.0cm]{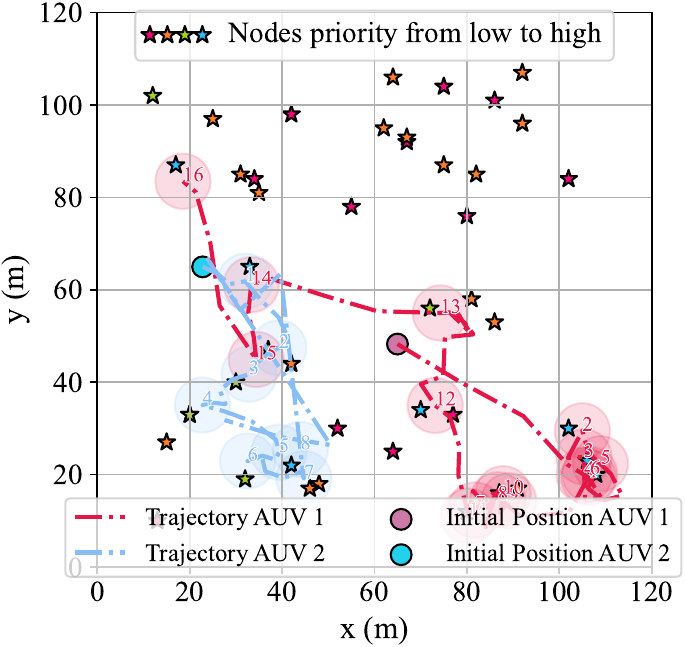}}
%  \vspace{1.5cm}
  \centerline{(b) AUV trajectories (MDP)}\medskip
\end{minipage}
\caption{The AUV trajectories using the expert policy trained via SAC algorithm based on AoI-MDP and standard MDP.}
\label{fig:1}
\end{figure}

Finally we guided the multi-AUV in the underwater data collection task using the expert policy trained via SAC algorithm based on AoI-MDP and standard MDP respectively. As illustrated in Fig. 5, the trajectory coverage trained under AoI-MDP is more extensive, leading to more effective completion of the data collection task. Conversely, under standard MDP, the trajectories of AUVs appears more erratic, with lower node coverage, thereby showcasing suboptimal performance.

\section{Conclusion}
In this study, we propose AoI-MDP to improve the performance of underwater tasks. AoI-MDP models observation delay as timing delay through SDF, and includes this delay as a new component in the state space. Additionally, AoI-MDP introduces wait time in the action space, and integrate AoI with reward functions to achieve joint optimization of information freshness and decision making for AUVs leveraging RL for training. Simulation results highlight the feasibility, effectiveness and generalization of AoI-MDP over standard MDP, which effectively minimizes AoI while showcasing superior performance in the underwater task. The simulation code has been released as open-source to facilitate future research.

\bibliographystyle{IEEEtran}
%\bibliography{refs}

\end{document}